# Analytical model for the approximation of hysteresis loop and its application to the scanning tunneling microscope

Rostislav V. Lapshin

*"Delta" Microelectronics and Nanotechnology Research Institute, 2 Schelkovskoye Shosse, Moscow 105122, Russia*

A new model description and type classification carried out on its base of a wide variety of practical hysteresis loops are suggested. An analysis of the loop approximating function was carried out; the parameters and characteristics of the model were defined – coersitivity, remanent polarization, value of hysteresis, spontaneous polarization, induced piezocoefficients, value of saturation, hysteresis losses of energy per cycle. It was shown that with piezomanipulators of certain hysteresis loop types, there is no difference in heat production. The harmonic linearization coefficients were calculated, and the harmonically linearized transfer function of a nonlinear hysteresis element was deduced. The hysteresis loop type was defined that possesses minimum phase shift. The average relative approximation error of the model has been evaluated as 1.5%–6% for real hysteresis loops. A procedure for definition of the model parameters by experimental data is introduced. Examples of using the results in a scan unit of a scanning tunneling microscope for compensation of raster distortion are given.

## I. INTRODUCTION

The law of a closed hysteresis curve would emerge in many physical phenomena such as dielectric hysteresis [polarization curve $P=f(E)$], magnetic hysteresis [magnetization curve $B=f(H)$], elastic hysteresis [deformation curve $\varepsilon=f(F)$], and some others.

This phenomenon is widespread and important, however a simple analytical equation that could approximate it with a sufficient degree of precision has not existed. Therefore, rather often when analyzing various processes and systems with a hysteresis element, the solution was being searched for either graphically – by using experimental data, or with the help of straight-line approximation of the curve.[1, 2]

The first method is inconvenient because of the need for table function representation and low precision of graphics. Low precision of the second method is caused by roughness of piecewise-linear approximation (certainly, if the number of the segments is not too great) and thereto it implies searching for the solutions at several intervals followed by "gluing" them with each other.

Other methods of hysteresis loop approximation are known in both polynomial models[2] and integral operators[3, 4] classes. But their usage is limited because of the complexity of hardware support or great time of calculations.

This work pursued two objectives. First, to give a description of the suggested model and its properties and characteristics. The part of the paper devoted to this matter is of a wide, general scope inasmuch as the use of the model is considered for analyzing static nonlinearity of hysteresis loops of various types and physical nature encountered in many scientific instruments. Second, to highlight the ways and manners of practical application of the conclusions obtained from the model description to particular purposes, namely, for description and compensation of nonidealities of piezoceramic manipulator[5, 6, 7, 8, 9, 10, 11, 12] of a scanning tunneling microscope (STM) – nonlinear-

## Analytical model for the approximation of hysteresis loop

Table I. Classification of hysteresis loops.

| No. | Type | Type status | $n$ | $m$ | $\beta$ | Equation | Figure |
|---|---|---|---|---|---|---|---|
| 1 | Leaf | Main | 1 | 1<br>3<br>5 | $\arctan\frac{b_y}{b_x}$ | $x(\alpha)=a\cos^m\alpha+b_x\sin^n\alpha,$<br>$y(\alpha)=b_y\sin\alpha$ | 1(a)-I<br>1(a)-II<br>1(a)-III |
| 2 | Crescent | Main | 2 | 1<br>3<br>5 | $\frac{\pi}{2}$ | or | 1(b)-I<br>1(b)-II<br>1(b)-III |
| 3 | Classical | Main | 3 | 1<br>3<br>5 | $\frac{\pi}{2}$ | $x(\alpha)=b_x\cos^n\alpha-a\sin^m\alpha,$<br>$y(\alpha)=b_y\cos\alpha,$<br>$0\le\alpha\le 2\pi$ | 1(c)-I<br>1(c)-II<br>1(c)-III |
| 4 | Tilted classical | Derivative from type 3 | a | a | $0\ldots\frac{\pi}{2}$ | $\bar{x}(\alpha)=x(\alpha)\cos\theta+y(\alpha)\sin\theta,$<br>$\bar{y}(\alpha)=-x(\alpha)\sin\theta+y(\alpha)\cos\theta,$<br>$\theta=(\pi/2)-\beta$ | 1(d) |
| 5 | Double loop | Derivative from types 1/3/4 | a | a | a | $\bar{\bar{x}}(\alpha)=x(\alpha)\pm b_x,$<br>$\bar{\bar{y}}(\alpha)=y(\alpha)\pm b_y$ | 1(e) |
| 6 | Bat | Derivative from types 1/3/4 | a | a | a | $\tilde{x}(\alpha)=x(\alpha),$<br>$\tilde{y}(\alpha)=\lvert y(\alpha)\rvert$ | 1(f) |

[a]Depending on the initial type.

ity and ambiguity of static characteristic, piezoceramic creep, thermal drift.

## II. DESCRIPTION OF THE MODEL

### A. Analytical expression for hysteresis curve family. Formation of the types and their classification

The family of hysteresis loops can be described by a generalized transcendental equation in parametric form as follows

$$\begin{aligned} x(\alpha) &= \pm a\cos^m\alpha \pm b_x\sin^n\alpha, \\ y(\alpha) &= b_y\sin\alpha, \end{aligned} \tag{1}$$

where $a$ is the split point coordinate; $b_x$, $b_y$ are the saturation point coordinates; $m$, $n$ are integer numbers (see Table I); $\alpha$ is a real parameter ($-\infty\le\alpha\le+\infty$).

Plus signs in Eq. (1) correspond to a hysteresis curve with saturation points in I, III quadrants (this is the case to be considered below) and minus signs in II, IV. Thus, the curves built with plus and minus signs would be symmetrical to each other relative to *oy* axis.

Classification of the hysteresis loop types supported by the model is presented in Table I. Each type is ascribed with a status of either main or derivative. The main types are “leaf,” “crescent” (see Ref. 13), “classical” [Figs. 1(a)-(c)]. They are determined by the $n$ coefficient. The $m$ and $n$ ($n$>3) powers in Eq. (1) define the steepness (I, II, III curves). The derivative types are “tilted classical,” “double loop,” “bat” [Figs. 1(d)-(f)]. They derive from main types or from other derivative types with implementing some extra operations.

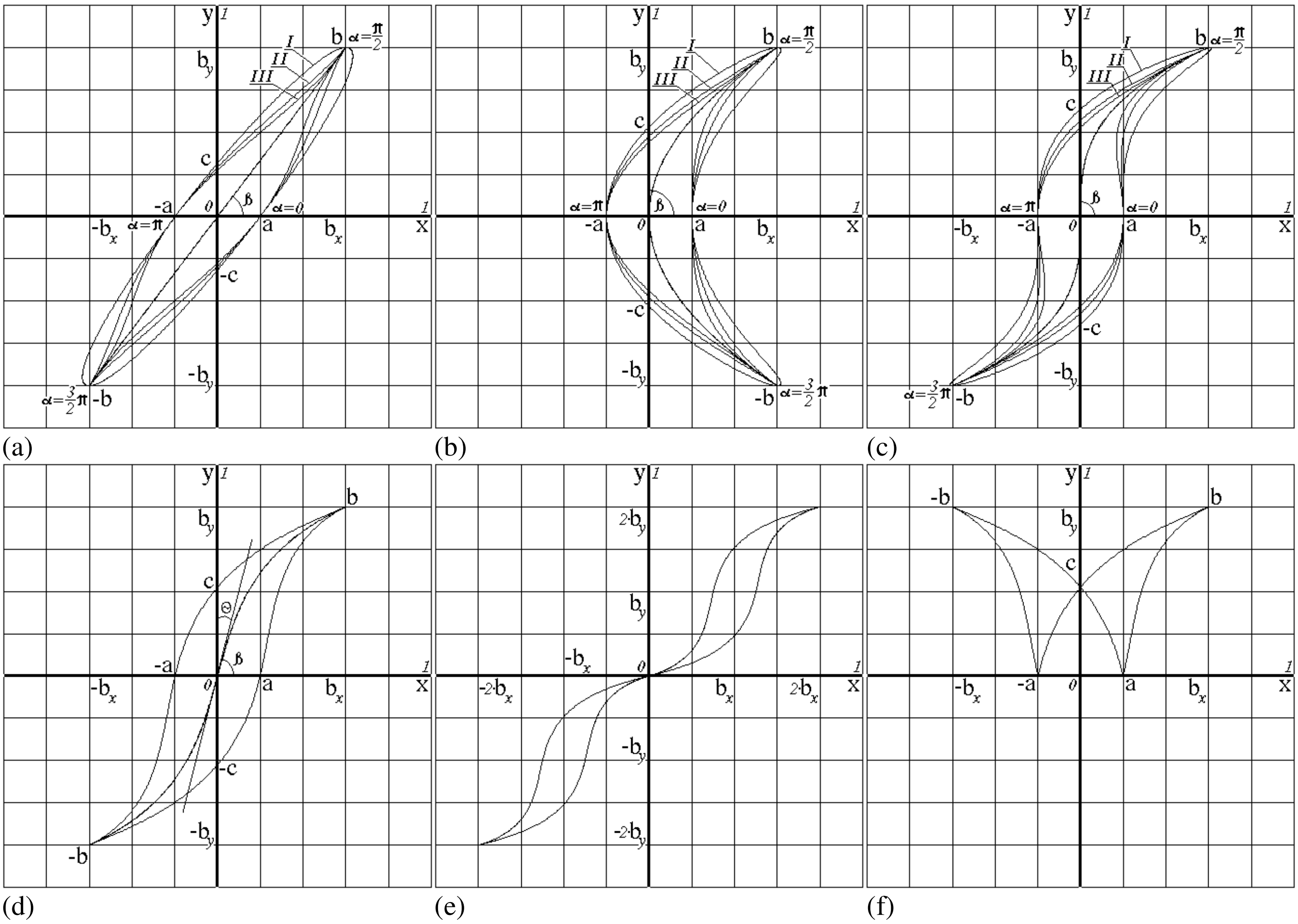

Fig. 1. Hysteresis loop types supported by the model: (a) Leaf ($a$=0.2; $b_x$=0.6; $b_y$=0.8; $m$=1, 3, 5; $n$=1; $\theta$=0°); (b) crescent ($a$=0.2; $b_x$=0.6; $b_y$=0.8; $m$=1, 3, 5; $n$=2; $\theta$=0°); (c) classical ($a$=0.2; $b_x$=0.6; $b_y$=0.8; $m$=1, 3, 5; $n$=3; $\theta$=0°); (d) tilted classical ($a$=0.2; $b_x$=0.6; $b_y$=0.8; $m$=3; $n$=3; $\theta$=15°); (e) double loop ($a$=0.1; $b_x$=0.4; $b_y$=0.4; $m$=3; $n$=3; $\theta$=15°); (f) bat ($a$=0.2; $b_x$=0.6; $b_y$=0.8; $m$=3; $n$=3; $\theta$=15°).

So, a curve of tilted classical derivative type [Fig. 1(d)], as the tangent in inflection point of the unsplit loop makes an angle $\beta\neq\pi/2$ with $ox$ axis, can be built by rotating the coordinate system clockwise through the angle $\theta=\pi/2-\beta$. Thus, using the foregone formula for transformation of Cartesian coordinates with the axes rotated,[14] the following equation is obtained:

$$\begin{aligned}\bar{x}(\alpha) &= x(\alpha)\cos\theta + y(\alpha)\sin\theta,\\ \bar{y}(\alpha) &= -x(\alpha)\sin\theta + y(\alpha)\cos\theta,\end{aligned} \tag{2}$$

where $\bar{x}(\alpha)$, $\bar{y}(\alpha)$ are the coordinates of the rotated system.

When rotating, the split point $\pm a$ and the saturation point $\pm b$ change their positions relative to the origin system. Therefore a preliminary distortion of their coordinates is needed to get these points to coincide with the originals after rotation. Here, the following transformation formulas[14] will be used:

$$\bar{a}_x = a\cos\theta \tag{3}$$

and

$$\begin{aligned}\bar{b}_x &= b_x\cos\theta - b_y\sin\theta,\\ \bar{b}_y &= b_x\sin\theta + b_y\cos\theta.\end{aligned} \tag{4}$$

## Analytical model for the approximation of hysteresis loop

As $a$, $b_x$, $b_y$ in Eq. (1) are substituted with the corrected value of split constant $\overline{a}_x$ and saturation constants $\overline{b}_x$, $\overline{b}_y$, equations for $x(\alpha)$ and $y(\alpha)$ can be obtained, which are needed in calculations by formula (2).

To build the models of more complicated hysteresis loops [e. g., double loops, Fig. 1(e), which are a composition of l/3/4 type] or to simulate thermal drift processes or creep effect, the rectangular coordinate system should be shifted by the value of $x_0$, $y_0$.[14] Herein, the value of $x_0$ is a bias voltage,[5] and for creep and thermal drift, the value of $y_0$ is a certain time function $y_0=f(t)$.[6, 7, 8, 9] Thus, the shifted curve could be described as follows:

$$\begin{aligned} \overline{\overline{x}}(\alpha) &= x(\alpha) \pm x_0, \\ \overline{\overline{y}}(\alpha) &= y(\alpha) \pm y_0, \end{aligned} \tag{5}$$

where the equations for $x(\alpha)$, $y(\alpha)$ are determined by the composition base type.

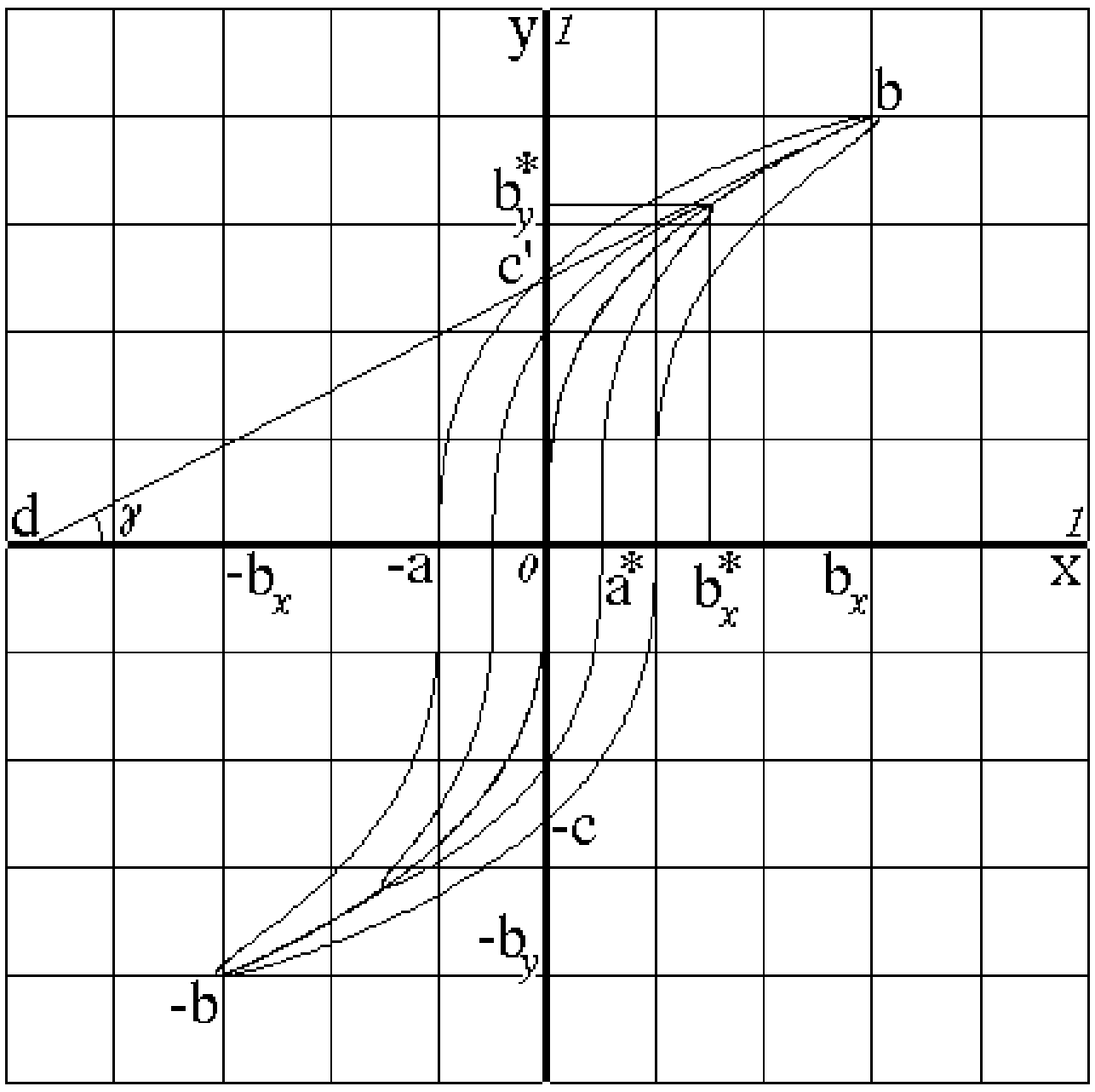

Fig. 2. Embedded loops. Determination of spontaneous polarization value ($a$=0.2; $b_x$=0.6; $b_y$=0.8; $m$=1; $n$=3; $b_x^*$=0.3).

To form a model of bat type [Fig. 1(f)], it is sufficient to take the module of $y(\alpha)$.

The direction of passing along a hysteresis loop is supposed to be determined in the following way.[1, 2] For a coordinate lag system, the movement along the lower half of the curve is associated with the inequality $dx/dt>0$, and along the upper half with $dx/dt<0$, the direction obtained being counterclockwise. For a coordinate lead system, the signs in those inequalities would be swapped (clockwise direction). When a parametrically set model is being considered, it is suitable to associate the positive (counterclockwise) direction with increasing $\alpha$ parameter, and the negative one with decreasing.

In the case of an unsplit hysteresis loop ($a$=0), it is easy to pass from the parametric record over to an explicit function definition

$$y(x) = \frac{b_y}{\sqrt[n]{b_x}} \sqrt[n]{x}, \quad -b_x \le x \le +b_x. \tag{6}$$

By using expression (6), loops embedded into the limit cycle can be built. The procedure consists of the following (see Fig. 2): for $b_x^*$ chosen within the ($-b_x$, $b_x$) interval, the value of $b_y^*$ is being found by Eq. (6), which lies at the unsplit curve. Then, the embedded loop being assumed, at first approximation, as similar to the limit cycle, the value of split $a^*=a\,b_x^*/b_x$ is being set and the curve is being built using the parameters $m$, $n$, $\theta$ of the limit cycle. In the general case, instead of Eq. (6), any suitable function that passes through the points $\pm b$ and the origin of coordinates can be used.

To trace the history of motion, it should be done in the following way. First, the moment is defined when the sign of derivative of the input signal is changing (this moment corresponds to a saturation point of a particular cycle). Next, as the value of the $\alpha$ parameter is known at that moment, the parameters $\pm b_x^*=x(\alpha)$, $\pm b_y^*=y(\alpha)$ can be calculated. Then, as the law of disposal of particular cycles inside limit one is known, it becomes possible to define $a^*$ and $\theta^*$ parameters. Finally, the $\alpha$ parameter is assigned the value of $\pi k/2$, where $k$=1, 3, 5, … Here are all the parameters needed to proceed with the motion along the new cycle.

Beginning with $m$=3 and $n$=1, function (1) cannot be resolved in explicit form $y=f(x)$ because it would result in an equation of more than fourth degree. However, an explicit record can be obtained for the function, inverse to Eq. (1), which is yielded by simply swapping $x(\alpha)$ and $y(\alpha)$

$$\begin{aligned} x(\alpha) &= b_y \sin\alpha, \\ y(\alpha) &= a\cos^m\alpha + b_x \sin^n\alpha. \end{aligned} \tag{7}$$

That inverse function will serve as the base for building a hysteresis compensation system. An explicit record for Eq. (7) can be given by

$$y = \frac{b_x}{b_y^n} x^n \pm \frac{a}{b_y^m} \sqrt{\left(b_y^2 - x^2\right)^m}, \quad -b_x \le x \le +b_x. \tag{8}$$

Taking into consideration, on one hand, that the operations of division and square root are not facile in hardware implementation, require much time of calculation, and are not typical in digital signal processor (DSP) algorithms and, on the other hand, the convenience of definition of the inverse function and the contour pass direction, the parametric form (1) must be admitted to be the most suitable to be used further.

## B. Analysis of the curve: Coersitivity, remanent polarization, dielectric spontaneous polarization, induced piezocoefficients, value of saturation

Function (1), which was taken as the model of hysteresis, is a continuous, nonlinear, ambiguous, limited one defined on the $[-b_x, b_x]$ segment. Besides, this function is periodical with the period $2\pi$, since the following equalities are true:

$$\begin{aligned} x(\alpha) &= x(\alpha + 2\pi k), \\ y(\alpha) &= y(\alpha + 2\pi k), \end{aligned} \tag{9}$$

where $k$=1, 2, 3, … Owning to the property of periodicity of function (1), use of the model will prove to be most effective with cycle processes, e. g., in scanning systems (see Sec. III.A).

Let us find the zeros of function (1). Here, the value of the $\alpha$ parameter should be defined such that $y(\alpha)$ becomes equal to 0, i. e., $b_y \sin\alpha=0$, whence $\alpha=\pi k$. Substituting this value of $\alpha$ in $x(\alpha)$, it will yield $x=\pm a$. Note that the physical sense of a zero of function (1) is coersitivity.

Now, the coordinates of the $\pm c$ point are to be defined as the intersection points of the loop and $oy$ axis. The point $\pm c$ defines the value of remanent polarization. Writing

$$\begin{aligned} a\cos^m\alpha + b_x \sin^n\alpha &= 0, \\ b_y \sin\alpha &= \pm c. \end{aligned} \tag{10}$$

In the case of $m=n$, this will result in the following solution:

$$\pm c = \frac{\pm b_y}{\sqrt{1 + \left(b_x/a\right)^{2/m}}}. \tag{11}$$

The value of hysteresis $H_y$ is defined as $c/b_y$(100%); by using expression (11), it can be written as

$$H_y = \frac{100\%}{\sqrt{1 + \left(b_x/a\right)^{2/m}}}. \tag{12}$$

Note, that the ratio $b_x/a$ in expression (12) is none other than $100\%/H_x$, where $H_x$ is "hysteresis" along $ox$ axis.

Thus, a correlation is obtained that links the values $H_y$ and $H_x$ with each other ($m$=$n$)

$$\left(\frac{100\%}{H_y}\right)^2-\left(\frac{100\%}{H_x}\right)^{2/m}=1. \tag{13}$$

Let us find the coordinate of the $c'$ point (see Fig. 2). It is the intersection point of the $oy$ axis and $bd$ tangent drawn through the saturation point of the unsplit curve (6). Note that the point $c'$ defines the value of dielectric spontaneous polarization. The equation for the $bd$ straight line is given by $y=kx+c'$, where $k=\tan\gamma$. The derivative $k=dy/dx$ of the unsplit curve (6) in the point $x=b_x$ is defined by $b_y/(nb_x)$. From the equation for the $bd$ line with $x=b_x$, $y=b_y$, taking into account the found $k$ value, the value of $c'$ is defined as

$$c'=b_y\left(1-\frac{1}{n}\right). \tag{14}$$

It will be shown (not rigorously) that function (1) is odd (except for crescent type). By definition, a function $y=f(x)$ would be odd provided $f(-x)=-f(x)$. In regard to function (1)

$$\begin{aligned}-x(\alpha)&=-\left(a\cos^m\alpha+b_x\sin^n\alpha\right),\\ -y(\alpha)&=-b_y\sin\alpha.\end{aligned} \tag{15}$$

Using the identities -sin $\alpha$=sin(-$\alpha$), cos $\alpha$=cos(-$\alpha$), making a formal replacement of the -$\alpha$ parameter with $\zeta$, and considering the oddness of $m$ and $n$ as well as the fact that the sign of the split parameter $a$ in formula (1) defines only the beginning point and the curve pass direction and, therefore, can be chosen as negative here, the following statements can be formulated:

$$\begin{aligned}-x(\zeta)&=a\cos^m\zeta+b_x\sin^n\zeta,\\ -y(\zeta)&=b_y\sin\zeta.\end{aligned} \tag{16}$$

Thus, the right-hand parts of the equations in (16) have no changes beside those in Eq. (1); consequently the function is odd. As the graph of an odd function is symmetrical relative to the origin of coordinates, the information of a half is sufficient to have the whole function set. The property of symmetry of function (1) gives the opportunity, for example at estimating the approximation error (see Sec. II.E), to carry out the calculations at a half of the domain.

Let us find first and second derivatives of function (1), which in case of a parametrically set function are given by[14]

$$\frac{dy}{dx}=\frac{\dot{y}(\alpha)}{\dot{x}(\alpha)} \tag{17}$$

and

$$\frac{d^2y}{dx^2}=\frac{\dot{x}(\alpha)\ddot{y}(\alpha)-\ddot{x}(\alpha)\dot{y}(\alpha)}{\dot{x}(\alpha)^3}, \tag{18}$$

where $\dot{x}(\alpha)=dx/d\alpha\neq0$, $\ddot{x}(\alpha)=d^2x/d\alpha^2$, $\dot{y}(\alpha)=dy/d\alpha$, $\ddot{y}(\alpha)=d^2y/d\alpha^2$ are first and second derivatives of the functions $x(\alpha)$ and $y(\alpha)$ by $\alpha$ parameter, respectively. For the function being considered

$$\dot{x}(\alpha)=-am\cos^m\alpha\tan\alpha+b_xn\sin^n\alpha\cot\alpha, \tag{19}$$

$$\ddot{x}(\alpha)=am\cos^m\alpha\left[(m-1)\tan^2\alpha-1\right]+b_xn\sin^n\alpha\left[(n-1)\cot^2\alpha-1\right], \tag{20}$$

$$\dot{y}(\alpha)=b_y\cos\alpha, \tag{21}$$

$$\ddot{y}(\alpha)=-b_y\sin\alpha. \tag{22}$$

In accordance with Eq. (17), the first derivative is defined as

$$\frac{dy}{dx}=\frac{b_y\cos\alpha}{-am\sin\alpha\cos^{m-1}\alpha+b_xn\cos\alpha\sin^{n-1}\alpha} \tag{23}$$

and, in accordance with Eq. (18) after simple transformations, the second derivative is defined as

$$\frac{d^2y}{dx^2}=\frac{ab_ym(2-m)\sin^2\alpha\cos^{m-1}\alpha+ab_ym\cos^{m+1}\alpha-b_xb_yn(n-1)\sin^{n+1}\alpha\cot^3\alpha}{\left(-am\sin\alpha\cos^{m-1}\alpha+b_xn\cos\alpha\sin^{n-1}\alpha\right)^3}. \tag{24}$$

Expression (23) allows for calculating the induced piezocoefficients of a hysteresis curve in the coordinates of displacement versus electric field strength as well as the differential magnetic permeance at any point of the curve $B=f(H)$.

By analyzing expressions (23) and (24), it can be shown that for $m=1$ function (1) reaches its maximum value in the point $(+b_x,+b_y)$ and minimum – in the point $(-b_x,-b_y)$. For all $m\neq1$, the lower and the upper parts of hysteresis loop (1) are monotonously increasing functions, therefore at the edges of the domain, i. e., in the points $\pm b_x$, function (1) has the greatest $+b_y$ and the least $-b_y$ values. Thus, the $(\pm b_x,\pm b_y)$ points of curve (1) are the saturation points of a hysteresis loop.

## C. Square of a loop

Physically, the square of a hysteresis loop characterizes the heat losses that cause heating of the material and, therefore, define its efficiency coefficient. To find the square $S$ of a loop, the following integral[14] should be calculated:

$$S=\frac{1}{2}\oint\left(x(\alpha)\frac{dy}{d\alpha}-y(\alpha)\frac{dx}{d\alpha}\right)d\alpha, \tag{25}$$

then taking into account expressions (1), (19), and (21), writing

$$S=\frac{1}{2}\int_0^{2\pi}\left[\left(a\cos^m\alpha+b_x\sin^n\alpha\right)b_y\cos\alpha-b_y\sin\alpha\left(-am\sin\alpha\cos^{m-1}\alpha+b_xn\cos\alpha\sin^{n-1}\alpha\right)\right]d\alpha. \tag{26}$$

Opening the parenthesis, grouping up the terms, then using De Moivre's expansion[14] of $\cos^{m+1}\alpha$ and $\cos^{m-1}\alpha$, and solving the integral for odd $n$, the hysteresis loop square (26) will be given by

$$S=\left[\frac{1}{2}C_{m+1}^{(m+1)/2}+m\left(C_{m-1}^{(m-1)/2}-C_{m-1}^{(m-3)/2}\right)\right]\frac{1}{2^m}\pi ab_y, \tag{27}$$

where $C_l^k=l!/[k!(l-k)!]$ are binomial coefficients. Thus, the square of leaf and classical hysteresis loops would not apparently depend on the saturation value by $x$ coordinate and is only determined by the split constant and the saturation value by $y$ coordinate.

Formula (27) is applicable to odd $n$, but the coefficient $n$ itself does not participate in it explicitly. Therefore, the following theorem can be formulated: the quantity of the heat produced by the piezomanipulator for a cycle is the same for hysteresis loops of both leaf and classical types, provided the values of $m$, $a$, $b_y$ involved are the same.

Since the square of a geometrical figure would be invariant to rotation, by substituting $a$ and $b_y$ in formula (27) with their corrected values $\bar{a}_x$ and $\bar{b}_y$ [see Eqs. (3), (4)], a formula can be written to define the square of tilted classical hysteresis loop

$$S=\left[\frac{1}{2}C_{m+1}^{(m+1)/2}+m\left(C_{m-1}^{(m-1)/2}-C_{m-1}^{(m-3)/2}\right)\right]\frac{1}{2^{m+1}}\pi a\left[b_x\sin2\theta+b_y(\cos2\theta+1)\right]. \tag{28}$$

**Analytical model for the approximation of hysteresis loop**

As the terms containing the $m$ coefficient are considered at various values of $m$ in formulas (27), (28), it can be concluded that with fixed values of the rest model parameters the inequality $S_{m=5}<S_{m=3}<S_{m=1}$ is true, which is seen well in Figs. 1(a), (c) (curves I, II, III). Thus, the greater the $m$ parameter is, the less heat the piezomanipulator produces.

For example, with $m=n=3$ ($\theta=0$) the losses of heat will be determined by the hysteresis loop square of $S=3/4\pi a b_y$. Note that with the product of $a$ by $b_y$ equal to $4/(3\pi)$, the hysteresis loop square becomes equal to unit.

## D. Harmonic linearization coefficients. The harmonically linearized transfer function of a hysteresis element

Let the linear part of a nonlinear hysteresis system have the property of a low-pass filter, i. e., transmit the first harmonic with practically no losses and considerably attenuate all the others, which have been produced by a nonlinear element under $x(\omega t)=A\sin(\omega t)$ input harmonic action. When analyzing such a system, the method of harmonic linearization can be applied, in accordance with which the hysteresis function $y=f(x)$ would be given in the following form:[1, 2]

$$y=\left(q(A)+\frac{\hat{q}(A)}{\omega}p\right)x, \tag{29}$$

where

$$\begin{aligned} q(A)&=\frac{1}{\pi A}\int_0^{2\pi} f(A\sin\alpha)\sin\alpha\, d\alpha,\\ \hat{q}(A)&=\frac{1}{\pi A}\int_0^{2\pi} f(A\sin\alpha)\cos\alpha\, d\alpha \end{aligned} \tag{30}$$

are the coefficients of harmonic linearization ($\alpha=\omega t$); $p$ denotes the differentiation operator ($p=d/dt$).

Let us define the harmonic linearization coefficients $q(A)$, $\hat{q}(A)$ for the inverse function (7). This function obviously matches the conditions of the task under consideration. At that, the input harmonic signal $x(\alpha)$ is given by $b_y\sin\alpha$, and $f(b_y\sin\alpha)$ – by $a\cos^m\alpha+b_x\sin^n\alpha$ (here, instead of $A$ amplitude, the denomination $b_y$, is involved), then in accordance with formula (30) writing

$$\begin{aligned} q(b_y)&=\frac{1}{\pi b_y}\int_0^{2\pi}\left(a\cos^m\alpha+b_x\sin^n\alpha\right)\sin\alpha\, d\alpha,\\ \hat{q}(b_y)&=\frac{1}{\pi b_y}\int_0^{2\pi}\left(a\cos^m\alpha+b_x\sin^n\alpha\right)\cos\alpha\, d\alpha. \end{aligned} \tag{31}$$

By solving the integrals in Eqs. (31) similarly to Eq. (26) for odd $n$, the coefficients of harmonic linearization of the inverse function (7) are given by

$$\begin{aligned} q(b_y)&=\frac{C_{n+1}^{(n+1)/2}b_x}{2^n b_y},\\ \hat{q}(b_y)&=\frac{C_{m+1}^{(m+1)/2}a}{2^m b_y}. \end{aligned} \tag{32}$$

In accordance with formula (29), the linearized equation for the hysteresis loop inverse function becomes the following:

$$y=\left(\frac{C_{n+1}^{(n+1)/2}b_x}{2^n b_y}+\frac{C_{m+1}^{(m+1)/2}a}{2^m b_y\omega}p\right)x, \tag{33}$$

and the transfer function $W^{1}(b_y, s)=Y(s)/X(s)$ becomes

$$W^{-1}(b_y,s)=q(b_y)+\frac{\hat{q}(b_y)}{\omega}s=\frac{C_{n+1}^{(n+1)/2}b_x}{2^n b_y}+\frac{C_{m+1}^{(m+1)/2}a}{2^m b_y\omega}s. \tag{34}$$

Now, the required harmonically linearized transfer function of a hysteresis element (1) is to be found as an inverse function to Eq. (34), i. e., $W(b_x, s)=1/W^{1}(b_y, s)$. Note that at that, the input amplitude will not be $b_y$ but $b_x$ instead. Finally

$$W(b_x,s)=\frac{2^{m+n}b_y\omega}{2^n C_{m+1}^{(m+1)/2}as+2^m C_{n+1}^{(n+1)/2}b_x\omega}. \tag{35}$$

By substituting $s=j\omega$ in Eq. (35) (where $j$ is imaginary unit), the amplitude-phase characteristic of a hysteresis element will be given by

$$W(b_x,j\omega)=W(b_x)=\frac{2^{2m+n}C_{n+1}^{(n+1)/2}b_x b_y-j2^{2n+m}C_{m+1}^{(m+1)/2}ab_y}{2^{2n}\left(C_{m+1}^{(m+1)/2}\right)^2a^2+2^{2m}\left(C_{n+1}^{(n+1)/2}\right)^2b_x^2}, \tag{36}$$

which only depends upon $b_x$ amplitude and does not upon $\omega$ frequency. Since $W(b_x)=q(b_x)+j\hat{q}(b_x)$, from Eq. (36) comes the next

$$\begin{aligned} q(b_x)&=\frac{2^{2m+n}C_{n+1}^{(n+1)/2}b_x b_y}{2^{2n}\left(C_{m+1}^{(m+1)/2}\right)^2a^2+2^{2m}\left(C_{n+1}^{(n+1)/2}\right)^2b_x^2},\\ \hat{q}(b_x)&=-\frac{2^{2n+m}C_{m+1}^{(m+1)/2}ab_y}{2^{2n}\left(C_{m+1}^{(m+1)/2}\right)^2a^2+2^{2m}\left(C_{n+1}^{(n+1)/2}\right)^2b_x^2}, \end{aligned} \tag{37}$$

which are the required harmonic linearization coefficients of function (1). For the case of $m=n=3$, coefficients (37) are given by $q(b_x)=4b_x b_y/\left[3\left(a^2+b_x^2\right)\right]$ and $\hat{q}(b_x)=-4ab_y/\left[3\left(a^2+b_x^2\right)\right]$; their graphs are shown in Fig. 3.

The coefficient $q(b_x)$ determines the steepness of the averaging line. From formula (37), it is easy to see that with the amplitude of input harmonic signal $b_x$ increasing, $q(b_x)$ approaches zero since in that case nonlinearity (1) gets saturated.

The amplitude of first harmonic $A(b_x)=|W(b_x)|$ can be defined by the transfer function (36) as

$$A(b_x)=\sqrt{q^2(b_x)+\hat{q}^2(b_x)}=\frac{2^{m+n}b_y}{\sqrt{2^{2n}\left(C_{m+1}^{(m+1)/2}\right)^2a^2+2^{2m}\left(C_{n+1}^{(n+1)/2}\right)^2b_x^2}} \tag{38}$$

and its phase $\varphi(b_x)=\arg W(b_x)$ – as

$$\varphi(b_x)=\arctan\frac{\hat{q}(b_x)}{q(b_x)}=-\arctan\frac{2^n C_{m+1}^{(m+1)/2}a}{2^m C_{n+1}^{(n+1)/2}b_x}. \tag{39}$$

For example, the phase shift of first harmonic at the output of classical nonlinear element shown in Fig. 1(c), curve II will make -18.4°.

The minus sign in expression (37) for $\hat{q}(b_x)$ [this coefficient is placed by the derivative in formula (29)] as well as the minus sign in Eq. (39) mean that the presence of a hysteresis element results in a phase lag of the output signal beside the input one. As is derived from Eq. (39), the wider the hysteresis loop is (the greater the split parameter $a$), the greater the phase shift becomes.

By analyzing expression (39), it can be shown that the following theorem would take place: with the same values of $m$, $a$, and $b_x$, minimum phase shift would occur at the leaf-type loops, and at that $|\varphi_{m=5}|<|\varphi_{m=3}|<|\varphi_{m=1}|$. Note here that it is the very type encountered in STM piezomanipulators [see Fig. 4(a) and Refs. 5, 9-12].

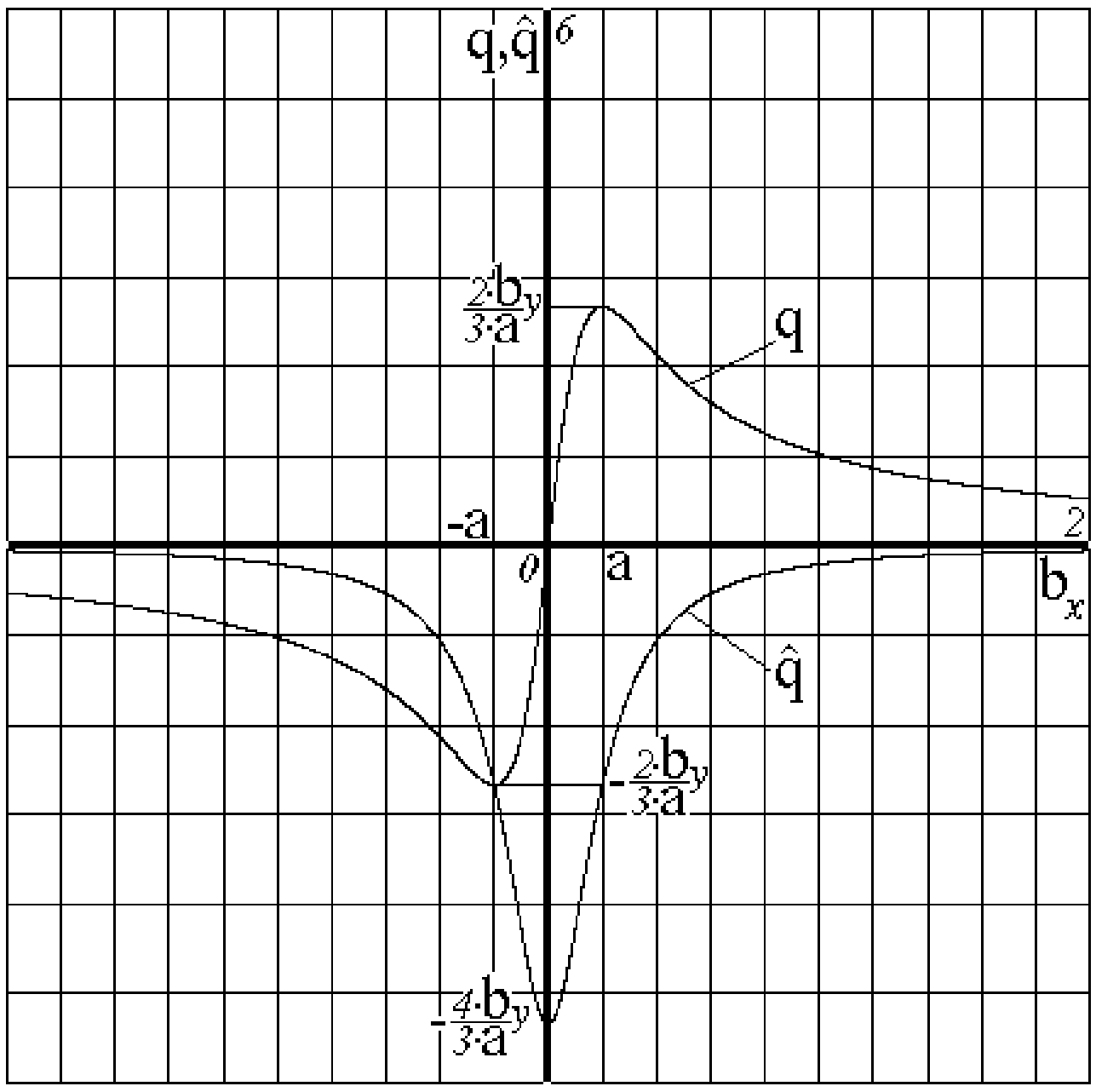

Fig. 3. Graphs of $q(b_x)$, $\hat{q}(b_x)$ harmonic linearization coefficients vs $b_x$ for the nonlinear hysteresis element ($a$=0.2; $b_y$=0.8; $m$=$n$=3).

The harmonically linearized transfer function (35) of a hysteresis element having been defined, the further analysis of an automatic control system should be carried out by employing the existing theory of linear systems in order to define its stability, precision, and quality of transient process.

## E. Estimates of approximation error

The error of approximating the experimental characteristics of hysteresis loops will be estimated by calculating the following quantities

(1) Maximal absolute approximation error

$$\Delta = \max_{0 \le x \le b_x} \left| y_m(x) - y_e(x) \right|, \tag{40}$$

where $y_m(x)$ are the model data; $y_e(x)$ are the experimental data.

(2) Maximal relative approximation error

$$\delta = \max_{0 \le x \le b_x} \frac{\left| y_m(x) - y_e(x) \right|}{\left| y_e^{\max} \right|} 100\% = \frac{\Delta}{\left| y_e^{\max} \right|} 100\%, \tag{41}$$

where $y_e^{\max} = b_y$ is the $y$ coordinate of the saturation point of experimental curve.

(3) Average relative approximation error

$$\langle \delta \rangle = \frac{100\%}{b_x \left| y_e^{\max} \right|} \int_0^{b_x} \left| y_m(x) - y_e(x) \right| dx. \tag{42}$$

(4) Quadratic mean approximation error ($\sigma$)

$$\sigma^2 = \frac{1}{b_x} \int_0^{b_x} \left( y_m(x) - y_e(x) \right)^2 dx. \tag{43}$$

The results of definition of the model approximation error of experimental hysteresis characteristics are presented in Table II, where the type and the parameters of a model are given.

## F. The procedure of definition of model parameters by limit cycle

For building the model, the six parameters must be extracted from the experimental dependence: $a$, $b_x$, $b_y$, $m$, $n$, $\theta$ by the following algorithm.

(1) By the shape of the curve, its type is being identified (see Table I). If the type proves to be derivative, then the main type is being identified, from which the derivative type has derived. When the type is known, the value of parameter $n$ and the kind of equation become known, also.

(2) If the type is identified as tilted classical or as its derivatives double loop or bat, then by the graph of experimental dependence, the angle $\theta$ is being defined.

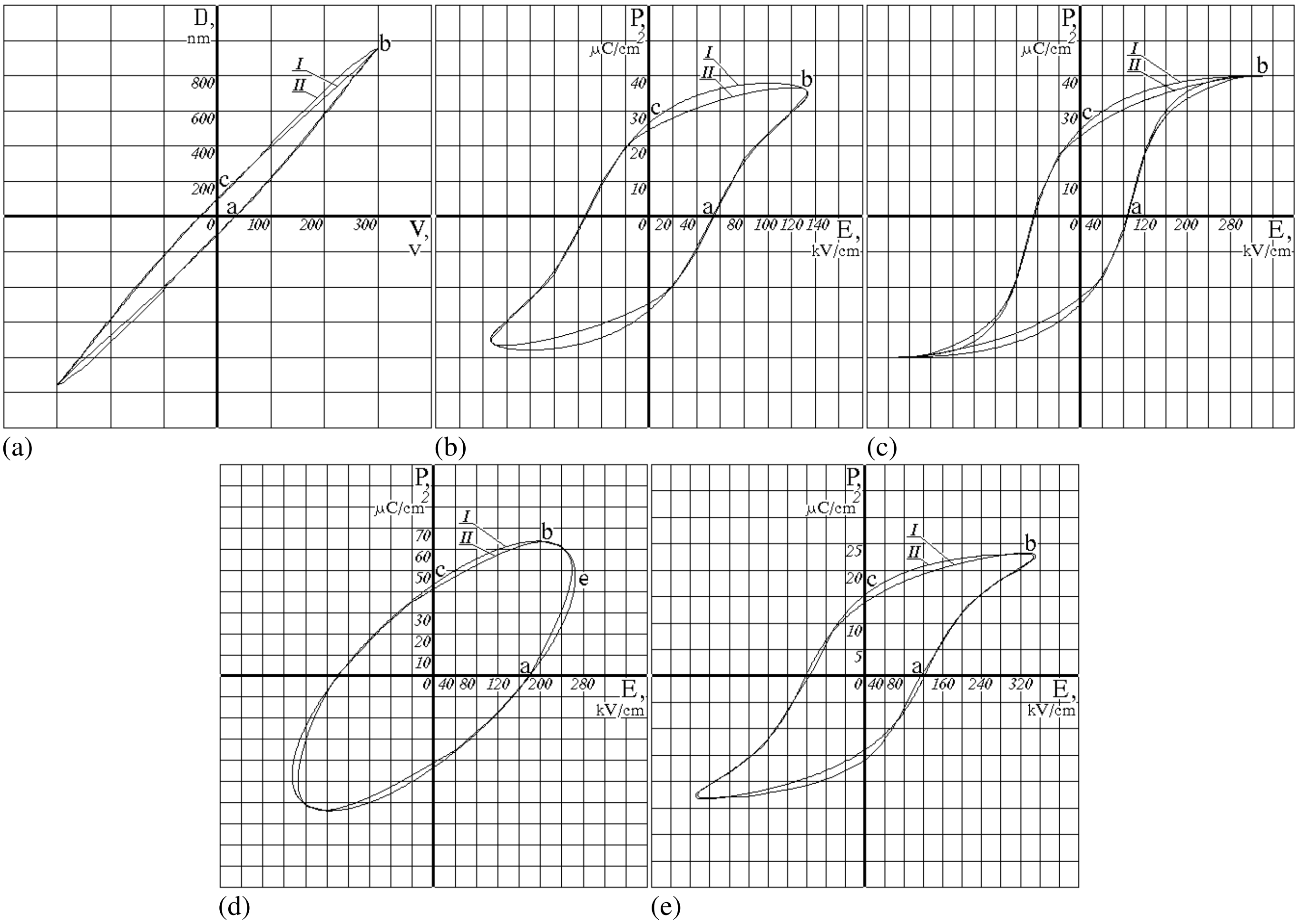

Fig. 4. Accuracy of approximation: I – model data; II – experimental data. (a) Displacement of STM PZT X, Y piezomanipulators vs applied voltage (leaf; $a$=32.6; $b_x$=300; $b_y$=955; $m$=3; $n$=1; $\beta$=72.6°; $H_y$≈11%; $\langle\delta\rangle$=1.5%); (b) ceramic polarization vs electric field strength (Ref. ) (tilted classical; $a$=54; $b_x$=130; $b_y$=36.4; $m$=1; $n$=3; $\theta$=27.7°; $\langle\delta\rangle$=6%); (c) ceramic polarization vs electric field strength (Ref. ); (tilted classical; $a$=90; $b_x$=340; $b_y$=39.6; $m$=5; $n$=3; $\theta$=30°; $\langle\delta\rangle$=4.1%); (d) ceramic polarization vs electric field strength (Ref. 16) (leaf; $a$=180; $b_x$=195; $b_y$=63.8; $m$=1; $n$=1; $\beta$=18.1°; $\langle\delta\rangle$=3.9%); (e) ceramic polarization vs electric field strength (Ref. 16) (tilted classical; $a$=122.5; $b_x$=341; $b_y$=23.2; $m$=1; $n$=3; $\theta$=14°; $\langle\delta\rangle$=2.7%).

(3) Provided that in a vicinity of the saturation point the curve behaves the way shown in Figs. 1(a)-(c) graph I, it means that $m$=1; and if like in graphs II, III (the same figure), then $m$>1.

(4) By the experimental dependence, the coersitivity value $a$ is being defined.

(5) By the experimental dependence, the $b_x$ and $b_y$ coordinates of the saturation point are being defined. If the model is of leaf type, then the angle $\beta$ can be calculated.

(6) If the model type is double loop, then the displacement values $b_x$, $b_y$ are being defined by the experimental dependencies.

(7) By varying the coefficients $m$ (if only $m\neq1$ – see item 3) and $n$ (if only $n\neq1$ – see item 1) and, within a small range, the value of $\theta$ angle (if only the type identified is not leaf or crescent) and using the estimates of Sec. II.E, the minimal value of approximation error is being achieved.

## III. APPLICATION OF THE MODEL

The hysteresis compensation elements (HCE), as they are inserted into the STM control system, are generally shown in Fig. 5. The compensation system described below relating to the class of open-loop systems, its HCE elements are serial nonlinear correcting units which could be realized in either hardware (in analog or digital embodiment) or software.

Table II. Results of calculation of the approximation error.

| Source of experimental data and figure | Type of model and its parameters: $a$; $b_x$; $b_y$; $m$; $n$; $\theta^o$ or $\beta^o$ | $\Delta$ | $\delta$(%) | $\langle\delta\rangle$ (%) | $\sigma$ |
|---|---|---|---|---|---|
| Fig. 4(a) | Leaf<br>32.6; 300; 955; 3; 1; 72.6 | 49.5 | 5.2 | 1.5 | 20.5 |
| Ref. 15, Fig. 4(b) | Tilted classical<br>54; 130; 36.4; 1; 3; 27.7 | 5.4 | 14.8 | 6.0 | 2.8 |
| Ref. 16, Fig. 4(c) | Tilted classical<br>90; 340; 39.6; 5; 3; 30 | 3.3 | 8.3 | 4.1 | 1.9 |
| Ref. 16, Fig. 4(d) | Leaf<br>180; 195; 63.8; 1; 1; 18.1 | 9.0 | 14.1 | 3.9 | 3.2 |
| Ref. 16, Fig. 4(e) | Tilted classical<br>122.5; 341; 23.2; 1; 3; 14 | 1.7 | 8.1 | 2.7 | 0.9 |

In order to compensate the hysteresis effect in all the three ways of realization, the function inverse to Eq. (1) will be employed so that to have the identity $f(f^{1}(X))\equiv X$ work (here, -1 designates an inverse function). In terms of the system being described, the source function $f$ is the piezomanipulator static characteristic approximated by Eq. (1), and the inverse function $f^{1}$ is function (7) synthesized by the HCE element. The inverse function argument $X$ is a control signal, for example "XIn" scan signal, which undergoes preliminary distortion in the HCE and then is applied, through a high voltage amplifier (HVA), to the piezomanipulator. Thus, the resulting displacement of the manipulator will correspond to the input function of the scan.

## A. Compensation of raster distortion in the STM scan unit

The flow chart of the hysteresis compensation unit is shown in Fig. 6(a). The scheme is built on analog elements and consists of the following units: GEN sinusoidal generator; a channel of MUL multipliers that carries out the operation of raising the input oscillation sin($\omega t$) to $m$ and $n$ powers; PS phase-shifting element that converts the $\sin^m(\omega t)$ oscillation into $\cos^m(\omega t)$ by shifting it by a quarter of the period; AMP2, AMP3 operational amplifiers with the gains $K=a$ and $K=b_x$, respectively; SUM summing amplifier; AMP1 amplifier with the gain $K=b_y$; COMP comparator, and AS analog storage unit. The last three components of the scheme are destined to set the accordance between the "XIn" input scan signal and the parameter $\alpha=\omega t$ by successive approximation [such embodiment is simpler than an explicit calculation of the function $\arcsin(X(t)/b_y)$].

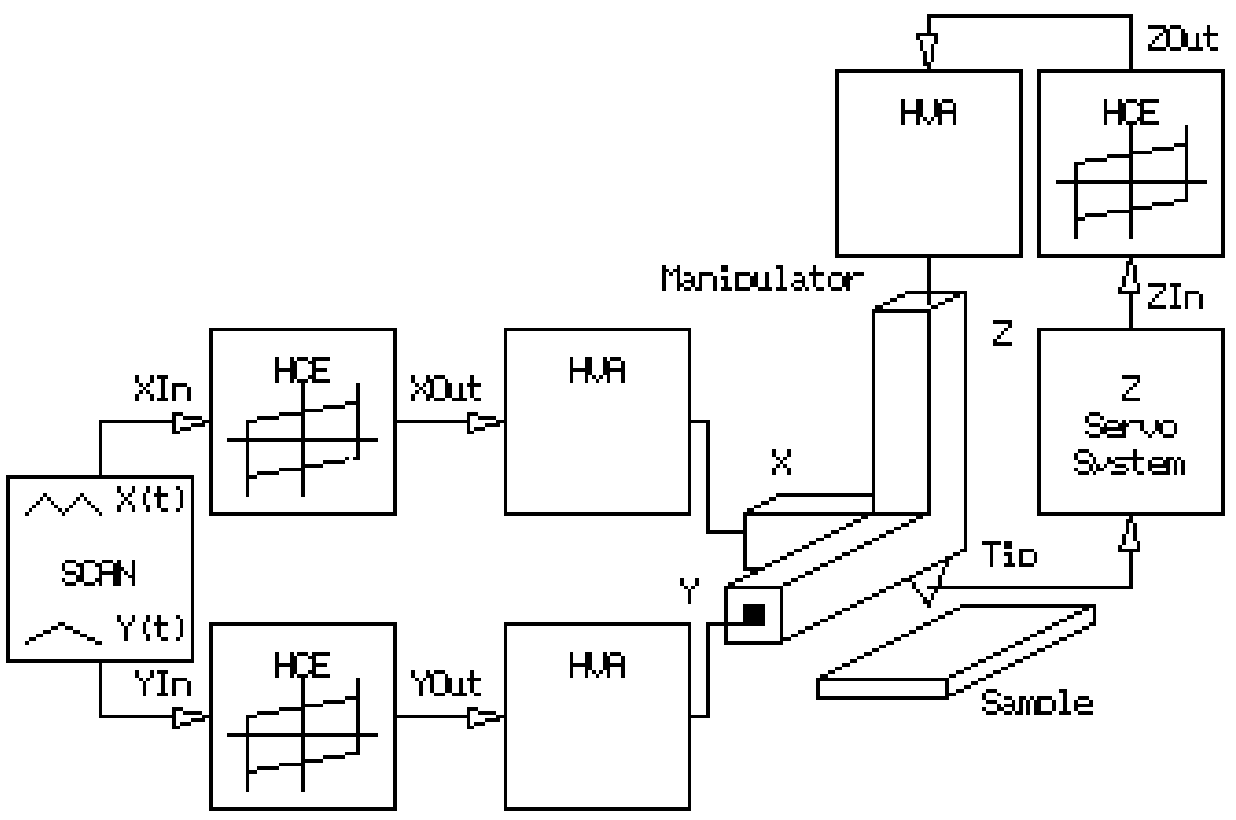

Fig. 5. HCE hysteresis compensation elements included in STM control system.

In fact, the scheme implements the calculation of the inverse function (7). Its operation principle is clear with no extra elucidation, to only point out that, to ensure the condition of in phaseness, the SCAN unit forming the scan works out an "XStart" signal at the beginning of each half-period of the scan, which starts the waiting generator GEN so that it would generate the proper half of the sine function accordingly to the sign of the scan signal derivative [the probing frequency of GEN is much greater than the frequency of the scan signal $X(t)$]. Pay attention to the triangle (not ramp) impulses of the scan signal shown in Fig. 6(a), the circumstance pointing out to operation with no idle stroke.

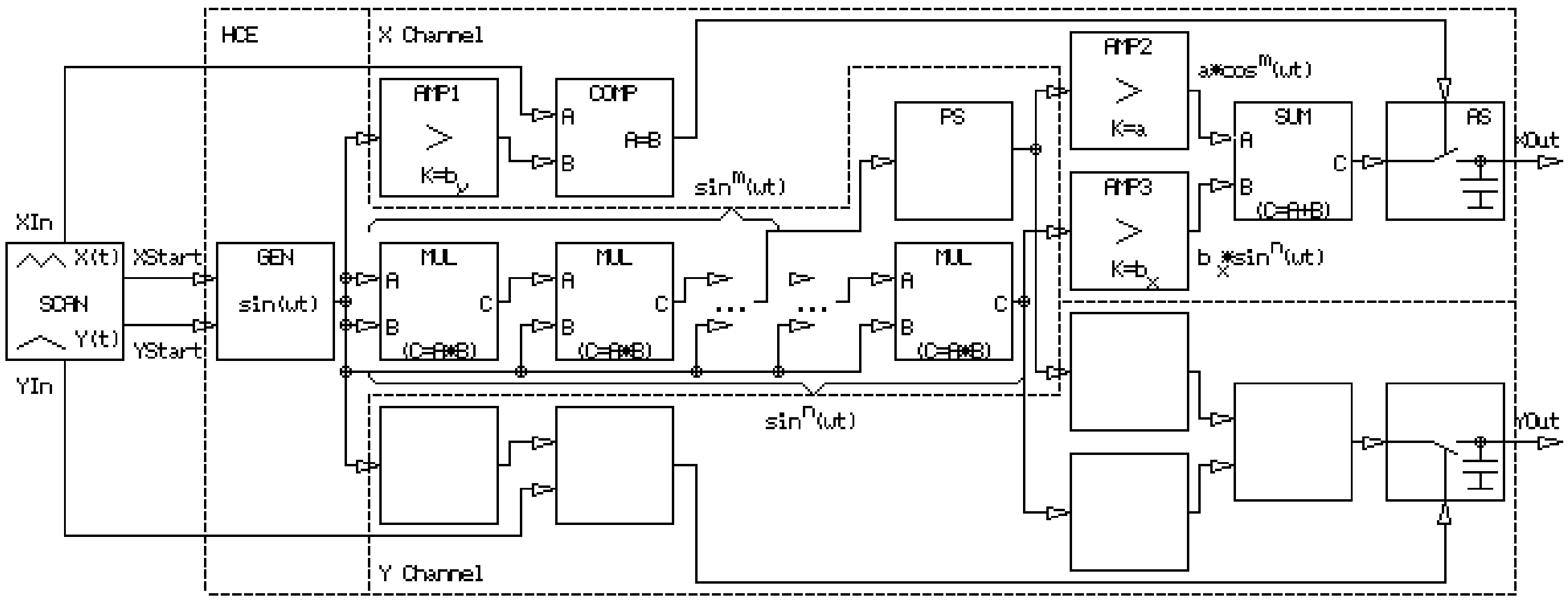

(a)

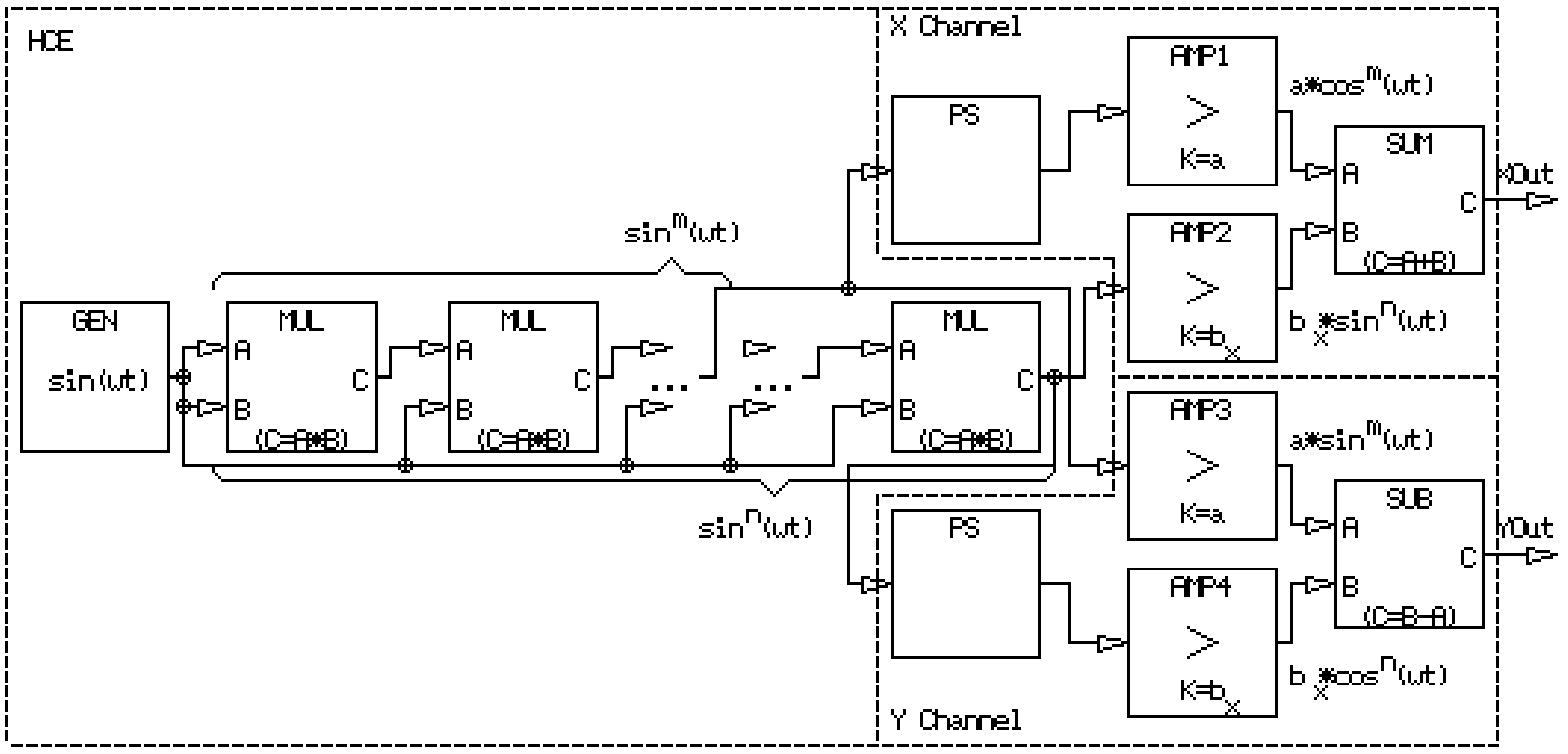

(b)

Fig. 6. Flow chart of analog device for hysteresis compensation for the cases of: (a) raster scan; (b) circular scan.

Let us show now what the scheme of the HCE element will look like in case of a circular scan microscope. Note that a circular scan is more preferable beside a raster scan for the following reasons: the design of X and Y manipulators proves to be symmetrical, an even employment of the manipulators occurs during the operation, the scan grows smoother and to those the possibility appears of getting the additional features of the microscope which have been described in paper 17. As a disadvantage of a circular scan, there can be admitted the necessity of converting the points of the circle into a conventional rectangular display window when visualizing.

To have a circular scan, it is necessary to apply a $\sin(\omega t)$ voltage to the X manipulator and $\cos(\omega t)$ to Y. Let us, instead of the triangle voltage [see Fig. 6(a)], mentally apply the $X(t)=b_y\sin\alpha$ voltage, where $\alpha=\omega t$. It can be easy to see that the chain of units – AMP1 amplifier, COMP comparator, and AS analog storage unit – becomes out of use and, therefore, can be excluded from the scheme. Now, instead of the triangle voltage $Y(t)$, let the voltage $Y(t)=b_y\cos\alpha$ be applied (structurally, X and Y channels are identical). Since a cosine function would outstrip a sine at π/2, formula (7) can be given by $y(\alpha)=a\cos^m(\alpha+\pi/2)+b_x\sin^n(\alpha+\pi/2)$, whence the inverse function can be written in another record

$$\begin{aligned} x(\alpha) &= b_y\cos\alpha, \\ y(\alpha) &= b_x\cos^n\alpha - a\sin^m\alpha, \end{aligned} \tag{44}$$

which can be used abreast with Eq. (7). From the last reasonings, it can be concluded that the circular scan generation scheme with simultaneous hysteresis compensation will take the outline presented in Fig. 6(b), where SUB is an operational amplifier connected in a differential mode. If comparing Figs. 6(a) and (b), the decrease of hardware expenditures can be revealed with the circular scan implementation; note that the parameter $b_y$ becomes unnecessary. The circular scan scheme described here would serve as the base for building cycloid and spiral scans.

The units of hysteresis compensation described above are the simplest ones and, generally speaking, do not permit us to work with loops of derivative types or various image sizes either to shift the image window along the scanning field or to implement vector access to a surface point. A hardware realization of the features mentioned above would result in a substantial complication of the equipment. The way out could be to build a digital system or to compute the model by a program.

A digital hysteresis compensation system will be designed by using the structures and the conclusions yielded while synthesizing the analog system. So, for the schemes presented in Figs. 6(a), (b), the analog multipliers MUL and the operational amplifiers AMP1/2/3/4 should be replaced with digital multipliers; the summing operational amplifier SUM and the differentiating amplifier SUB – with an arithmetic-logic unit; the analog storage unit AS – with a strobed register with a digital-to-analog converter (DAC) connected up to its output; the sinusoidal generator GEN – with a read-only memory (ROM) scheme with a sine table written down in it; the analog comparator COMP – with a digital one. A ring counter should be connected to the ROM address inputs. The function of the phase-shifting element PS consists of shifting the value of the address worked out by the ring counter so as to skip exactly a quarter of the period of sine in the ROM table.

At program realization of the model, the functions $\cos^m\alpha$, $\sin^n\alpha$ are stored in the computer data memory as some table structures. The scan voltage can be transformed into the corresponding $\alpha$ value by either successive approximation (similar to the operation of the unit chain AMPl-COMP-AS) or immediately by an arcsin table also kept in the data memory, or by directly calculating the arcsin by the foregone identity[14]

$$\arcsin\frac{f(t)}{b_y} = \int_0^t \frac{df(t)}{\sqrt{b_y^2 - f(t)^2}},$$

where $f(t)$ is a scan function ($|f(t)|<b_y$). At a half-period of the triangle scan, $f(t)$ is given by $kt$. So, by passing from the integral over to a sum, the following expression for $\alpha$ is obtained

$$\alpha = k\Delta t\sum_{i=0}^{N-1}\frac{1}{\sqrt{b_y^2 - k^2\Delta t^2 i^2}}, \tag{45}$$

where $N$ is the number of the samples taken; $\Delta t=t/N$ is the time discrete.

The calculation of the model could be fulfilled at a universal processor but most effectively this task could be resolved at a DSP because the operation $AB_i+C$ (where $A$, $B$, $C$ are real numbers) is encountered in the calculations which is typical of DSP algorithms. Here, $A=a$, $B_i=\cos^m\alpha_i$, $C=DE_j+F$, and $D=b_x$, $E_j=\sin^n\alpha_j$, $F=0$. The indexes $i$ and $j$ refer to the addresses of the memory cells where the tabulated values of $\cos^m \alpha_i$ and $\sin^n \alpha_j$ are contained, respectively.

The flow chart of a hysteresis compensation digital unit is shown in Fig. 7. This unit is realized on a soft-hardware base and ensures working with all the hysteresis types and performing a scan along an arbitrary trajectory so as to support any scan type, to carry out rotation of the scan window around *oz* axis (in order, e. g., to reduce moire distortion) as well as to vary the window size and shift it within the scan field.

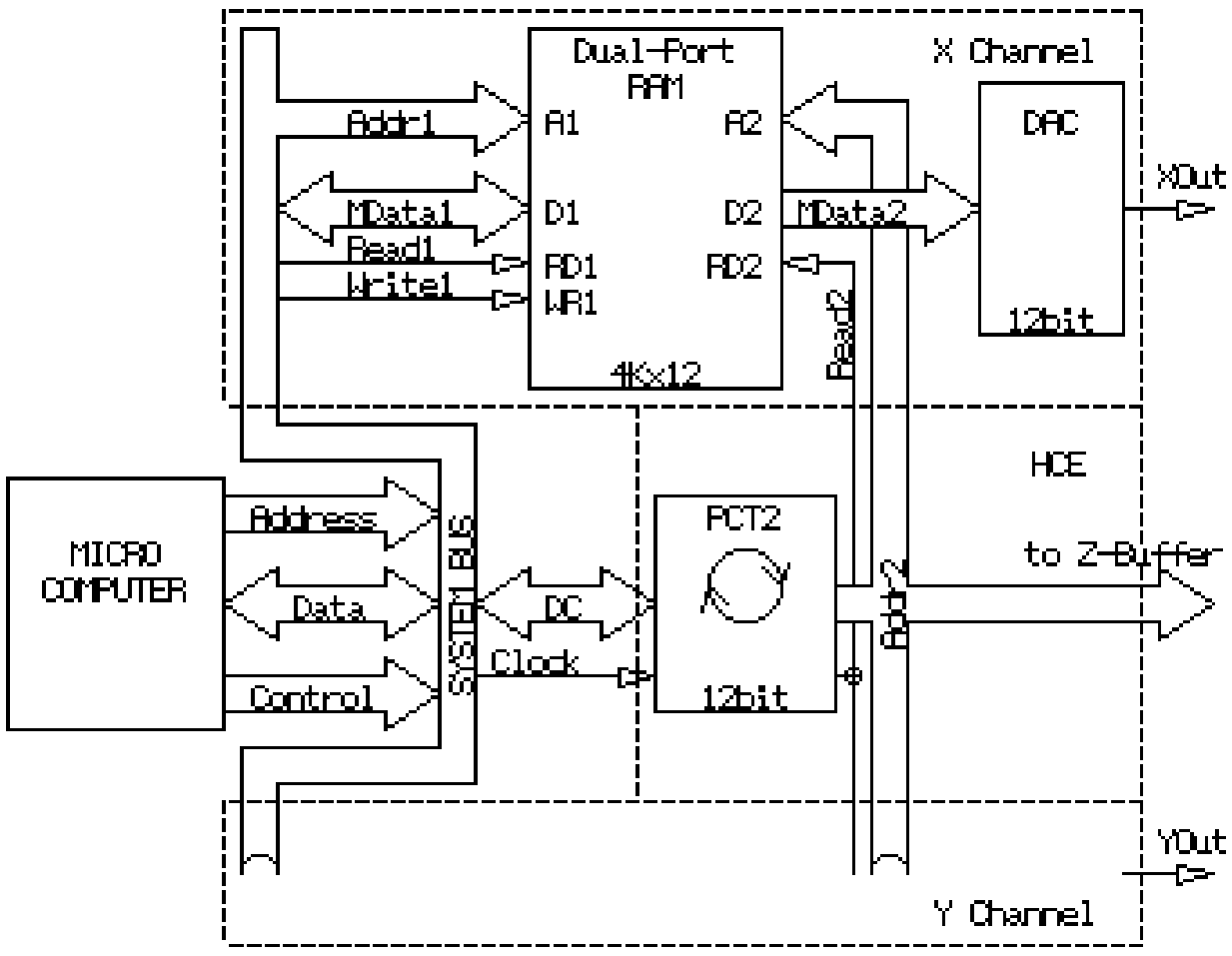

Fig. 7. Flow chart of digital device for hysteresis compensation (soft-hardware realization).

A ring programmable counter PCT2 is involved in the scheme which, in the rate defined by the “Clock” signal and the division coefficient “DC,” generates the address “Addr2” and the read signal “Read2” for the Dual-Port RAM (random-access memory) scheme with the model data written in it. From the RAM output, the “MData2” code, which corresponds to the preliminary distorted current value of raster voltage, comes into the DAC from where the signal, after having passed through a high voltage amplifier, is applied to the manipulator. Another RAM port is intended for writing the model data “MData1” calculated by the microcomputer processor. The use of a dual-port RAM permits us to get the processes of calculating the model and generating the control signal for the manipulator coincided in time and, therefore, to increase the unit fast acting.

Built on a dual-port RAM base, the scan subsystem is capable of being transformed from synchronous into asynchronous. To do that, it is sufficient, instead of the “Clock” signal, to apply a signal pointing out to readiness of the data in the tunnel junction stabilization system.[18] Besides, if that system also uses a dual-port RAM as a Z-Buffer, then for the next memory cell sampling, the “Addr2” signal generated by the PCT2 counter can be used. Thus, there can be seen a good mutual coordination in work between the tunnel junction stabilization system and the scan signal generation system.

It is appropriate to note that when using the scan modes that require frequent changing the model parameters *a*, $b_x$, $b_y$ (e. g., in order to change the image size), in a certain operation time, some error will have been accumulated in the model. At the moment when it reaches a certain threshold defined by the admissible approximation error, a compulsory correction must be done: for the model it is assigning $\pi/2$ to the $\alpha$ parameter and for the piezoelement – applying the saturation voltage to it so as to ensure it set at a fixed point of the limit characteristic – the saturation point.

## IV. EXPERIMENTAL RESULTS

The surface image of a test pattern is shown in Figs. 8(a), (c), which was taken by a STM when scanning with idle stroke and without, accordingly. As the test pattern, a diffraction grating with 0.3 μm period coated with gold was used. Since the grating structure changes along one direction only and scan piezomanipulators, because of hysteresis, introduce distortions along two directions, then for reflecting these distortions on the image taken, the test object was fitted up so that its stripes would make an angle of some 45° with the manipulator X axis.

On the images, the distortions of the test pattern are seen well, which consist in curving the stripes and chang-ing their width from one stripe to another [Fig. 8(a), cf. Ref. 11], splitting the stripes and formation of a double-sided comb-like structure [Fig. 8(c)], as well as parallel stripes looking like divergent ones [Fig. 8(e), cf. Ref. 12].

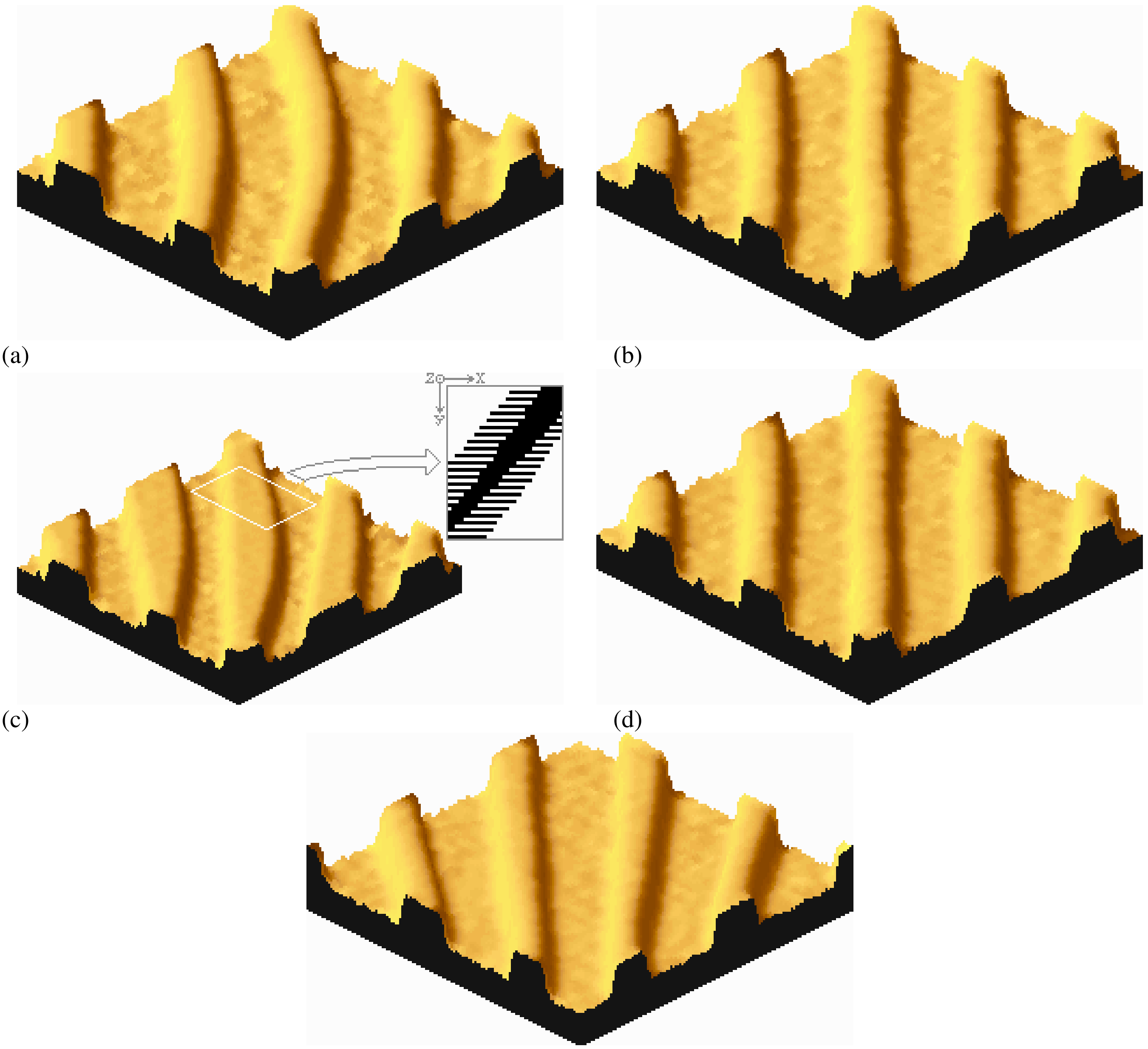

(a) (b) (c) (d) (e)

Fig. 8. 128×128 STM scan of the same 1×1 μm$^2$ area of test pattern surface (diffraction grating with 0.3 μm period coated with gold) (a), (c) Surface image data obtained by scanning with and without idle stroke, accordingly (the test object has been rotated by 45° counterclockwise), the hysteresis compensation system turned off. (b), (d) Surface image data corresponding to (a), (c) obtained with the hysteresis compensation system turned on. (e) Surface image data obtained by scanning with idle stroke (the test object has been rotated by 45° clockwise). After compensation of hysteresis, the image will look like the one shown in (b).

Figs. 8(b), (d) present the corrected images of the same area that were taken using the hysteresis compensation system shown in Fig. 7 [the model parameters were extracted from the data presented in Fig. 4(a)]. A visual comparison of Figs. 8(a) and (b) [(c) and (d)] will show that nonlinear distortions caused by hysteresis of manipulator piezoceramic are practically removed.

Hysteresis loops with parameter $m$ equaling to unit would have a section of negative derivative [see Fig. 4(d)]. The behavior of piezomanipulator at that part of the curve is similar to creep: on achieving $e$ point the voltage begins to decrease, the displacement still increasing for a certain time until the point $b$ is reached. [The parameter $\alpha$ that corresponds to $e$ point can be found by setting the denominator in Eq. (23) equal to zero.] From works 7, 8, it is known that creep would cause distortions at the image boundaries where the tip motion is reversed (e. g., straight lines would get hooks at the ends). Distortions of the same kind would appear when a piezomanipulator with the hysteresis loop said above is applied. Thus, the resulting picture will contain distortions of true creep and hysteresis mixed up together.

## V. DISCUSSION

The approximating model suggested belongs to the real time model class. It allows for compensation of the STM piezomanipulator nonidealities such as nonlinearity and ambiguity, which makes it possible to get rid of the image distortions (see Fig. 8 and Refs. 10-12), which is especially important with great scans, to refuse the idle stroke in the raster scan, and consequently to reduce the scanning time; in addition, the model supports the mode of direct access to a desired point. Therein, the reconstruction of the true image is available for either preprocessing (preliminary distortion of the scan trajectory) or post-processing of the distorted image taken.

As was shown in Sec. II.C, the $Q$ quantity of the heat produced by the piezomanipulator for a hysteresis cycle $t$ determined by the system timer or by the scan frequency can be easily calculated, for instance, by formula (28). The values $Q$ and $t$ might serve as the input data for some model considering the heat exchange proceeding among the STM construction elements. After having defined the difference in the manipulator temperature $\Delta T$ by using such model, the difference in the manipulator length $\Delta l$ can be found by the formula

$$\Delta l = \beta l \Delta T, \tag{46}$$

where $\beta$ is the heat expansion coefficient (1/K); $l$ is the initial manipulator length (m). Calculated in such a way, the thermal drift $\Delta l$ can be compensated in the scan unit by properly shifting the scan window. Thermal drift compensation is especially important in a scan device since there is not any active servo system mounted in it. Note that the STM heat processes, depending much upon construction features of the microscope and the materials used,[9] are of a rather complicated nature and deserve a special investigation which comes out of the limits of this paper.

The main advantage of the model, its simplicity, is called forth by no need of calculating the parameters – they are taken right from the experimental curve, neither the inverse function is to be calculated. To that, the model parameters make clear physical sense and have a simple geometrical interpretation. A distinctive feature of the model is the possibility of comparing different hysteresis loop types with each other (see theorems in Secs. II.C, II.D).

As a drawback of the model, there could be mentioned a restriction on the approximation accuracy to be achieved, i. e., its dependence upon the particular shape of the hysteresis curve. Though, the reverse task could be apparently resolved, the piezoceramic hysteresis curve being adopted to the desired model by changing the technological parameters, namely, the chemical composition, the baking and cooling conditions, the mechanical influences, etc.

Owing to application of the model, it becomes possible to linearize the STM tunnel junction stabilization system contour by methods similar to those described in Sec. III.A, which allows us to prevent the appearance of auto-oscillations and to get rid of distortions in Z direction,[8] although the method that was used in Ref. 18 must be admitted as the most acceptable solution here.

When measuring the frequency characteristic of Z manipulators, an auxiliary piezoelement is often employed, which is supposed to modulate the tunnel junction at a harmonic law. With this method, if a sine voltage is applied to the auxiliary piezoelement, the law of its mechanical displacement would not be exactly the sine because of the presence of hysteresis. The model for approximation of hysteresis loops would allow to increase the precision of this method. To do that, the voltage applied to the auxiliary piezoelement must match $x(\alpha)$ by formula (1). In that case, the displacement $y(\alpha)$ will be a harmonica1 function.

The model described may prove useful with the tasks of imitation modeling as well as in engineer calculations of nonlinear control systems containing hysteresis elements.

## ACKNOWLEDGMENTS

I want to thank Oleg E. Lyapin, Valery V. Efremov, Oleg D. Cnab, Vladimir N. Yakovlev, and Oleg V. Obyedkov for their helpful advice and discussions.

## APPENDIX

Beside smooth loops, the model suggested can be implemented for description of piecewise-linear loops as well (see Refs. 1, 2). To obtain a set of piecewise-linear hysteresis loop primitives, in formula (1) there must be used some piecewise-linear functions instead of sin $\alpha$ and cos $\alpha$. They could be, for instance, trapeziumlike pulses with unit amplitude

$$\begin{aligned} \mathrm{trp}_s(\alpha) &= \sum_{i=0}^{\infty}\left(\frac{4}{D-d}\left(\alpha - i\frac{T}{2}\right)(-1)^i \,\mathrm{rect}_1(\alpha,i) + (-1)^i \,\mathrm{rect}_2(\alpha,i)\right), \\ \mathrm{trp}_c(\alpha) &= \mathrm{trp}_s\left(\alpha + \frac{T}{4}\right), \end{aligned} \tag{A1}$$

where the subscripts $s$ and $c$ refer to sine and cosine, respectively; $d$ and $D$ are the upper and lower bases of the trapezium, respectively; $T=d+D$ is the period of pulses; $\mathrm{rect}_1(\alpha, i)=1(\alpha+(D-d)/4-iT/2)-1(\alpha-(D-d)/4-iT/2)$ and $\mathrm{rect}_2(\alpha, i)=1(\alpha-(D-d)/4-iT/2)-1(\alpha-(D-d)/4-d-iT/2)$ are the $i$th rectangular pulses; $1(\alpha, i)$ is the $i$th unit step function. Triangle functions $[\mathrm{tri}_s(\alpha)=\lim_{d\to 0}\mathrm{trp}_s(\alpha)$, $\mathrm{tri}_c(\alpha)=\mathrm{tri}_s(\alpha+T/4)]$ or rectangle functions $[\mathrm{rect}_s(\alpha)=\lim_{d\to D}\mathrm{trp}_s(\alpha)$, $\mathrm{rect}_c(\alpha)=\mathrm{rect}_s(\alpha+T/4)]$ could be also used to that purpose.